\documentclass[aps,prl,twocolumn, showpacs]{revtex4}

\usepackage{epsfig}
\usepackage{graphicx}
\def\be{\begin{equation}}
\def\ee{\end{equation}}
\def\bea{\begin{eqnarray}}
\def\eea{\end{eqnarray}}

\begin{document}
\title{1D Exciton Spectroscopy of Semiconductor Nanorods}
\author{A. Shabaev and Al. L. Efros$^*$}
\affiliation{Center for Computational Material Science, 
Naval Research Laboratory, Washington DC
20375}

\begin{abstract}
We have theoretically shown that optical properties of semiconductor nanorods are controlled by 1D excitons. The theory, which takes into account anisotropy of spacial and dielectric confinement, describes size dependence of interband optical transitions, exciton binding energies. We have demonstrated that the fine structure of the ground exciton state explains the linear polarization of photoluminescence. Our results are in good agreement with the measurements in CdSe nanorods.
\end{abstract}

\maketitle

There is growing interest in nano-size crystalline semiconductor structures of various shapes such as nanocrystals (NCs) \cite{Ekimov}, nanorods (NRs) \cite{NR1} and nanowires (NWs) \cite{Lieber} created by the "from the bottom up" technological approach.  Size-tunable control of their optical and transport properties combined with the ability  to move these nano-size objects around with precise control opens the exciting possibilities for the creation of new functional materials which can be used in unlimited applications. Among these nanostructures, the NCs are the most heavily studied and one can find a broad description of their properties and their potential applications in the reviews of Brus\cite {Brus}  and Alivisatos\cite{Alivisatos}. 

The optical properties of NRs, however, differ significantly from those of NCs. Compared to NCs, the NRs show higher photoluminescence (PL) quantum efficiency  \cite{NR1},  strongly linear polarized PL\cite{NRpolarization,NRPeng}, an increase of the global Stokes shift \cite{NRpolarization}, and significantly faster carrier relaxation \cite{NREl-Syed}. The Auger processes in NRs are strongly suppressed relative to those in NCs \cite{NRKlimovAPL,NRKlimovPRL},  which subsequently decreases the optical pumping threshold for stimulated emission \cite{NRKlimovPRL,NRlasing}. The size and shape dependence of the optical and tunneling gaps measured in CdSe NRs \cite{NRBanin} shows an unexpectedly large difference that cannot be explained by the electron-hole Coulomb correction to the optical gap used for NCs. To describe their measurements Katz {\em et al.} \cite{NRBanin} applied the four-band theory developed by Sercel and Vahala \cite{SV90}. The most important result of the latter paper was the prediction of an inverse order of light and heavy hole subbands in NWs, which was later confirmed by semiempirical pseudopotential calculations \cite{NRpolarization,NRGapCalc,JPC} and numerical calculations within the 6 band model \cite{NR6bandCalc}. An anisotropy of the light hole-to-electron optical transition matrix element predicts some degree of the PL linear polarization parallel to the NW \cite{SV91,Gershoni}. 

The dielectric confinement connected with the difference between dielectric constants of semiconductor crystallites, $\kappa_s$, and surrounding medium, $\kappa_m$, (for example, see \cite{Keldysh}) usually does not affect the optical spectra of NCs because the charge distributions of an excited electron and hole practically compensate each other at each point of the NC. This leads to a complete screening of an electric field of the total charge from penetration into the surrounding medium. This is not the case, however, in NRs where the electron and hole are at a distance larger than a NR radius and interact predominantly through the surrounding medium, which usually has a small dielectric constant, $\kappa_m\ll\kappa_s$. This  should lead to a formation of 1D excitons (1DEs) with large binding energy similar to that in NWs \cite{NWMulyarov}. In addition, the  dielectric confinement of anisotropic crystallites leads to strong PL polarization and PL polarization memory that was observed in NWs \cite{Jwang} and in porous Si \cite{p-Si}. 
This dielectric model, however, does not explain the strong linear polarization of NRs \cite{NRpolarization}.

In this letter, we have developed a theory that describes energy spectra and polarization properties of 1DEs in NRs. Our theory takes into account both spatial and dielectric confinement and allows one to describe optical properties of elongated NRs with quasi-cubic and zinc-blende lattice structure. The calculation for CdSe NRs surrounded by a dielectric with $\kappa_m=2.0$ has shown that the 1DEs control  optical properties of NRs even at room temperatures and has described the difference between the optical and tunneling gaps \cite{NRBanin}. We have shown that the fine structure of 1DE
explains the observed 87\% polarization degree of PL \cite{NRpolarization} and leads to the fast radiative decay in NRs at room temperatures.  

In what follows, we consider the NR as a crystalline that has the shape of an ellipsoid of revolution with major semi-axis $b$ (the NR axis) much larger than  the minor semi-axis $a$ (the NR radius). The motion of electrons and holes is analyzed within the 6-band model that has successfully described optical spectra of CdSe NCs \cite{JOSA}. The significant elongation of NRs ($b\gg a$) makes it possible, by using an adiabatic approximation, to separate the parallel motion from the motion perpendicular to the NR axis. Within this approximation we initially neglect the slow motion of carriers parallel to the NR axis. Assuming that the momentum of the parallel motion $k_z\approx 0$, we find the spectrum of carriers in 2D confinement perpendicular to the NR axis. Next, we consider the parallel motion by averaging the Hamiltonian over the fast motion of carriers strongly confined in 2D. 

Each electron state in NRs is characterized by the angular momentum projection on the NR axis, $m=0,\pm 1,...$.  The wave functions of the electron motion perpendicular to the NR axis can be written as $\sim e^{im\phi}J_{|m|}( k\rho)$, where $\phi$ is the azimuthal angle, $\rho$ is the distance from the NR axis, and $J_l(x)$ is the Bessel function of the $l$-th order. The momentum $\hbar k$ is connected to the kinetic energy of the electron motion perpendicular to the NR, $E_e=\hbar^2k^2/2m(E_e)$, where $m(E)=m_0/\{ 1 + 2f + (E_p/3)[ 2/(E + E_g) + 1/(E + E_g + \Delta)]\}$ is the electron effective mass that depends on the electron energy, $E$,  $m_0$ is the free electron mass, $E_g$ is the band gap, $\Delta$ is the spin-orbit splitting of the valence band, $E_p$ is the Kane energy parameter, and the parameter $f$ includes the contribution of remote bands. 
Assuming that the wave function vanishes at the NR surface, we find the momentum $\hbar k$ which satisfies the boundary condition: $\hbar k_{n|m|}= \hbar\alpha_{n|m|}/a$, where $\alpha_{n|m|}$ is the $n$-th root of the Bessel function of the $|m|$-th order. This gives the following energy of electron subbands $E_{n|m|}$ with the corresponding wave functions $\Psi_{n|m|}^{(e)}(\rho,\phi)$:
\begin{equation}
E_{n|m|}={\hbar^2 \alpha_{n|m|}^2\over 2a^2m(E_{n|m|})},~\Psi_{nl}^{(e)} = 
{J_{|m|}( \alpha_{n|m|}\bar{\rho})e^{im\phi}\over
\sqrt{\pi}a |J_{|m|}^{\prime}( \alpha_{n|m|})|},
\end{equation}
where $\bar{\rho} = \rho/a$ and the three smallest $\alpha_{n|m|}$ are $\alpha_{10}\approx 2.41$,  $\alpha_{11}\approx 3.83$, $\alpha_{12}\approx 5.14$.

Due to a strong spin-orbit coupling in the valence band, each hole state in the NR is characterized by the total angular momentum projection on the NR axis
$j_z=m+J_z$ which is the sum of angular momentum projection and the projection of the hole spin $J=3/2$ ($j_z=\pm 1/2,\pm3/2,...$). The energy spectrum of holes is found within the 6-band model, which takes into account the nonparabolicity of light hole spectrum \cite{ARMS}. In this model, the hole wave function of the motion, perpendicular to the NR axis, with total angular momentum projection, $j_z$, can be written
\bea
&\Psi_{j_z}^t(\mbox{\boldmath$r$})&=\sum_{\mu=-3/2}^{3/2}C^{3/2,t}_\mu J_{|j_z-\mu|}(x_t\bar{\rho}){e^{i(j_z-\mu)\phi}\over \sqrt{2\pi}}u^v_{3/2,\mu}\nonumber\\
&+&\sum_{\mu=-1/2}^{1/2}C^{1/2,t}_\mu J_{|j_z-\mu|}(x_t\bar{\rho}){e^{i(j_z-\mu)\phi}\over \sqrt{2\pi}}u^v_{1/2,\mu}
\eea
where $u^v_{1/2,\mu}$ and $u^v_{3/2,\mu}$ are the valence band Bloch functions \cite{ARMS}. For every value of  the hole energy $E$, there are dimensionless momenta $x_t$ and the sets of coefficients $C^{\nu,t}_\mu$ corresponding to the three branches in the dispersion law of the valence band ($t=l,h,s$): 
\bea
x^2_h&=&{\varepsilon\over \gamma^L_1-2\gamma^L},~x^2_{l(s)}={\eta(\varepsilon)+(-)\xi(\varepsilon)
\over (\gamma^L_1-2\gamma^L)(\gamma^L_1+4\gamma^L)},~\nonumber\\
\eta(\varepsilon)&=&2\varepsilon(\gamma^L_1+\gamma^L)-\delta(\gamma^L_1+2\gamma^L),\nonumber\\
\xi(\varepsilon)&=&\sqrt{\eta(\varepsilon)^2-4\varepsilon(\varepsilon-\delta)(\gamma_1-2\gamma)(\gamma^L_1+4\gamma^L)},
\eea
where $\varepsilon=E/E_0$, $\delta=\Delta/E_0$, $E_0=\hbar^2/2m_0a^2$, and the energy dependent Luttinger parameters $\gamma^L_1$ and $\gamma^L$ are connected with the contributions from remote bands, $\gamma_1$ and $\gamma$, by the relationships: $\gamma^L_1=\gamma_1+E_p/[3(E_g+E)]$ and  $\gamma^L=\gamma+E_p/[6(E_g+E)]$ \cite{ARMS}. At $k_z=0$, the 6-band Hamiltonian can be decomposed into two 3-band Hamiltonians and the wave functions  $\Psi_{j_z}^{t,\pm}(\mbox{\boldmath$\rho$})$ in Eq.(2) are described by the set of only three coefficients: $\mbox{\boldmath$C$}^{t,\pm} \equiv (C^{3/2,t}_{\pm 1/2},C^{3/2,t}_{\mp 3/2},C^{1/2,t}_{\pm 1/2})$. For the heavy hole and for the mixture of the light-- and spin-orbit split holes, these coefficients are $\mbox{\boldmath$C$}^{h,\pm} =(\sqrt{3},1,0)$ and $\mbox{\boldmath$C$}^{s(l),\pm} =(1,-\sqrt{3}, \chi_{s(l)}/i\sqrt{2})$ respectively,
with 
$\chi_{s(l)} = 1  - \left[\delta/2 +(-) \sqrt{\delta^2/4 -\gamma^L \delta x_{s(l)}^2 + (3\gamma^L x_{s(l)}^2)^2}\right]/ (\gamma^L x_{s(l)}^2)$.

\begin{figure}[th]
\vskip-0.7truecm
\begin{center}
\epsfig{file=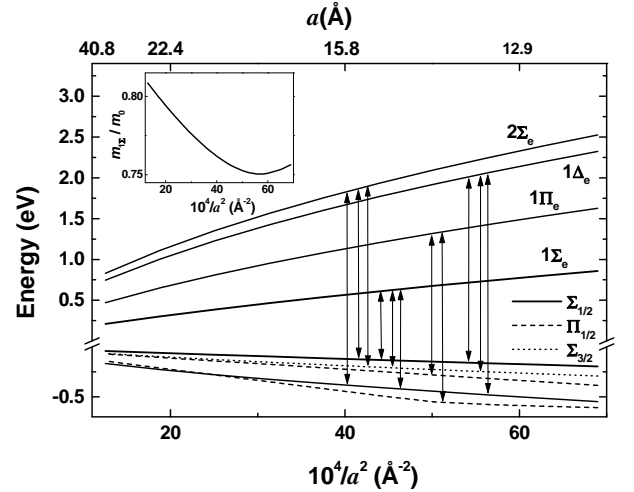, angle=-90, width=0.5\textwidth}
\end{center}
\vskip-0.5truecm
\caption{Size dependence of the electron and hole energy spectrum in CdSe NRs ($E_g=1.839$\,eV, $E_p=19$\,eV, $\Delta=0.42$\,eV, $f=-1.035$, $\gamma=0.55$, and $\gamma_1=2.1$). Arrows show optically allowed interband transitions. 
         Inset: Size dependence of the hole effective mass at  the $1\Sigma_{1/2}$ subband.}
\vskip-0.1truecm
\label{2Dlevels}
\end{figure}

The hole wave function of the motion perpendicular to the NR axis is a linear superposition of the three eigenfunctions $\Psi_{j_z}^{h,\pm}$, $\Psi_{j_z}^{l,\pm}$, and $\Psi_{j_z}^{s,\pm}$, corresponding to the same energy (see Eq.(3)). The requirement that the wave function vanishes at the NR surface defines the coefficients of this linear superposition and gives the dispersion equation for the hole quantum size levels (QSLs). The wave function  can be written:
\be
 \Psi_{j_z}^{\pm}( \rho,\phi) = 
 C^\pm_{j_z} [C_{j_zh}^\pm \Psi_{j_z}^{h,\pm} + C_{j_zs}^\pm \Psi_{j_z}^{s,\pm} + \Psi_{j_z}^{l,\pm}]~,
\ee
where $C_{j_zh}^\pm =  J_{|j_z\mp1/2|}(x_l)(\chi_l/\chi_s-1)/[\sqrt{3}J_{|j_z\mp1/2|}(x_h)]$,
$C_{j_zs}^\pm = -\chi_l J_{|j_z\mp1/2|}(x_l)/[\chi_s J_{|j_z\mp1/2|}(x_s)]$, and the coefficient $C^\pm_{j_z}$ is determined by the normalization condition: $\int d^2\rho |\Psi_{j_z}^{\pm}( \rho,\phi)|^2=1$.
The energy of the hole QSLs are given by the solutions of:
\bea
   3g_{j_z}^\pm(x_s, x_h, x_l) &-&3 g_{j_z}^\pm(x_l, x_h, x_s)+ g_{j_z}^\pm(x_h, x_s, x_h)\nonumber\\
    &-& g_{j_z}^\pm(x_h, x_l, x_h) = 0~,
\eea
where $g_{j_z}^\pm (u, v, w ) = v^2 \left[ u^2 - (\gamma^L_1 + 2\gamma^L)w^2/(\gamma^L_1 - 2\gamma^L)\right] 
\times J_{|j_z\pm3/2|}\left(u \right)J_{|j_z\mp1/2|}\left(v \right)J_{|j_z\mp1/2|}\left(w \right)$.

The size dependence of the electron and hole QSLs calculated for CdSe NRs is shown in Fig.1, where we use level notations adopted from molecular physics.
For the electron levels it is $n\Lambda_e$, where $\Lambda=\Sigma,\Pi,\Delta,...$ for the states with angular momentum projection on the NR axis $|m|=0,1,2,...$. For the hole it is $n\Lambda_{|j_z|}$, where $\Lambda$ corresponds to the smaller value of $|m|$ and $|m+2|$, the absolute values of the two angular momentum projections represented in the hole wave function, and $n$ is the number of the level for the given symmetry. The
6 band model calculation shows that the  $1\Sigma_{1/2}$ hole level with $|j_z|=1/2$ is  above the $1\Sigma_{3/2}$ level with $|j_z|=3/2$ in consistence with previous predictions of the reverse order of the light and heavy hole QSLs in NRs. The arrows in Fig.1 show the optically allowed transitions between the electron and hole QSLs, which have been obtained by the calculation of the overlap integral $K_{n\Lambda,n'\Lambda'_{|j_z|}}=\int d^2\rho \Psi_{nl}^{(e)}(\mbox{\boldmath$\rho$})\Psi_{j_z}^{\pm}(\mbox{\boldmath$\rho$})$ (see Ref.\cite{JOSA}). 
The energy shift $\delta \varepsilon_{n\Lambda_{j_z}}$ of each 1D hole subband 
at finite $k_z$  can be described in terms of the 1D effective mass: 
$\delta \varepsilon_{n\Lambda_{j_z}}(k_z) = \hbar^2 k_z^2/2m_{n\Lambda_{j_z}}$.
The dependence of the first hole subband effective mass, $m_{1\Sigma_{1/2}}$, on the NR radius  is shown in the inset of Fig.1.

The calculated optical spectra, however, does not include the electron-hole ({\em e-h}) Coulomb interaction enhanced by the penetration of an electric field into a surrounding medium. These {\em e-h} interaction can be written
\bea
 U \left( \mbox{\boldmath$r$}_e,\, \mbox{\boldmath$r$}_h \right)& =&  -e^2/\left(\kappa_s\left|\mbox{\boldmath$r$}_e - \mbox{\boldmath$r$}_h \right|\right)
  - e V_s\left(\mbox{\boldmath$r$}_e,\, \mbox{\boldmath$r$} _h \right)\nonumber\\
 &+& e V_s\left(\mbox{\boldmath$r$}_e,\, \mbox{\boldmath$r$}_e \right)/2
  + eV_s\left(\mbox{\boldmath$r$}_h,\, \mbox{\boldmath$r$}_h \right)/2~,
\eea
where the first term is the direct {\em e-h} interaction in a semiconductor, the second term
is the {\em e-h} interaction mediated by the surface separating the NR from the surrounding
dielectric, and the last two terms are self-interactions of each particle with the surface.
For an ellipsoid, the potential $V_s$ can be written in the  following form \cite{Smythe} 
\bea
 &&V_s\left(\mbox{\boldmath$r$},\, \mbox{\boldmath$r$}^{\prime} \right) ={e\over \kappa_s } 
  \sum_{n=0}^{\infty}\sum_{m=0}^{n}
\frac{( -1)^m(2 - \delta_{m0})(2n+1) S_n^m}
{\sqrt{b^2 - a^2}[(n+m)!/(n-m)!]^2}
   \nonumber\\ 
 &&\times   P_n^m\left(\eta^{\prime} \right)P_n^m\left(\eta \right)
   P_{n}^{m}\left( \xi^{\prime} \right) P_n^m\left(\xi \right)\cos m\left(\phi^{\prime} - \phi \right)~,
\eea
where $P_n^m$, $Q_n^m$, $\dot{P}_n^m$, and $\dot{Q}_n^m$ are Legendre functions \cite{SpFunct}
of spheroidal coordinates ($\eta$  and $\xi$) and their derivatives, $\delta_{ij}$ is the Kronecker symbol, and
\be 
   S_n^m =    
    {\left( \kappa_m - \kappa_s \right)Q_n^m\left(\eta_s\right)\dot{Q}_n^m\left(\eta_s\right)\over
    \kappa_s \dot{P}_n^m\left(\eta_s\right)Q_n^m\left(\eta_s\right) 
    - \kappa_m \dot{Q}_n^m\left(\eta_s\right)P_n^m\left(\eta_s\right)}.
\ee
The spheroidal coordinates are connected with cylindrical coordinates ($z$ and $\rho$) by the relationships: $\eta \xi \sqrt{b^2 - a^2}  = z$ and
$\left[\left(1 - \xi^2 \right)\left(\eta^2 - 1 \right)\left(b^2 - a^2 \right)\right]^{1/2} = \rho$. At the spheroidal surface of the NR $\eta$ becomes  $\eta_s = b/\sqrt{b^2 - a^2}$.

In the adiabatic approach, one averages the potential $U \left( \mbox{\boldmath$r$}_e,\, \mbox{\boldmath$r$}_h \right)$ in Eq.6 over the electron and hole wave functions of the fast motion perpendicular to the NR axis (Eq.1 and Eq.4).  The procedure leads to the 1D potential of {\em e-h}  Coulomb interaction and the shift of the electron and hole QSLs due to their interaction with the "mirror forces"  described by the last two terms in Eq.6. The 1D potential $U_{1\Sigma_e-1\Sigma_{1/2}}(z)$ for the lowest $1\Sigma_e$ electron  and $1\Sigma_{1/2}$ hole  subbands   can be approximated as 
\be
  U_{1\Sigma_e-1\Sigma_{1/2}}(z)\approx U_{\rm eff}\left( z\right) = -e^2/[\kappa_m(|z| + \rho_{\rm eff})]
\ee
where $\rho_{\rm eff} = 0.7 a$ for $\kappa_s=6.1$. This approximation is valid only if the distance $\left|z \right| = \left|z_e - z_h \right|$
between two charges is much smaller than the NR length $2b$. 

In the 1D adiabatic potential, 1DEs are formed under each pair of {\em e-h} subbands.  These 1DEs strongly modify the absorption and PL spectra of NRs. 
The approximation of the 1D potential in Eq. 9 makes it possible to analytically find energy spectra and wave functions of 1DEs. The wave function of the 1DE is 
$\Phi_{\rm ex}\left(z \right) = A W_{\alpha, 1/2}\left(\bar{z} \right)$, where $W_{\alpha, 1/2}$ is the Whittaker function \cite{SpFunct}, $A$ is the normalization constant, and $\bar{z} = \pm 2\left(\left|z\right|+  \rho_{\rm eff} \right)/\left(a_{1D}\alpha \right)$ with
$a_{1D} = e^2\kappa_m/\hbar^2\mu$, where $\mu=[1/m(E)+1/m_{n\Lambda_{j_z}}]^{-1}$ is the reduced electron and hole effective mass.  The plus sign is for positive $z$ and the minus sign is for negative $z$. Each state of the 1DE is characterized by its parity. For "even" states, including the ground state, the derivative of the wave function
must turn to zero at $z=0$. The wave function of "odd" states becomes zero itself at $z = 0$.
The boundary conditions determine the $n$-th value of $\alpha_n$ and the energy spectrum of 1DEs as $E_n = - \hbar^2/(2\mu a_{1D}^2\alpha^2_n )$. 

The size dependence of the CdSe NR transport gap, which is determined by the energy between the lowest $1\Sigma_e$ electron and $1\Sigma_{1/2}$ hole subbands, is shown in Fig.2 for the bare QSLs (dotted line) and for the QSLs shifted by the "mirror forces"  (dashed line). One can see that the "mirror forces" corrections to the transport gap of NRs are  significant. The optical gap in NRs is determined by the 1DEs. The size dependence of the first two "even" 1DE states (only "even" states are optically active)  is shown in Fig.2 by solid lines.  Even in thick NRs, the binding energy of the ground and excited states is higher than 150\,meV and 50 meV, respectively.  As a result, 1DEs should completely control the absorption and PL of CdSe NRs even at room temperature.
Given the good agreement with the experimental data for the transport and optical gaps in CdSe NRs \cite{NRBanin}, formation of the 1DEs in NRs explains the substantial difference ($\sim240-300$\,meV) between the two gaps in NRs with $a=1.8-2.0$\,nm.

\begin{figure}[th]
\vskip-1.2truecm
\begin{center}
\epsfig{file=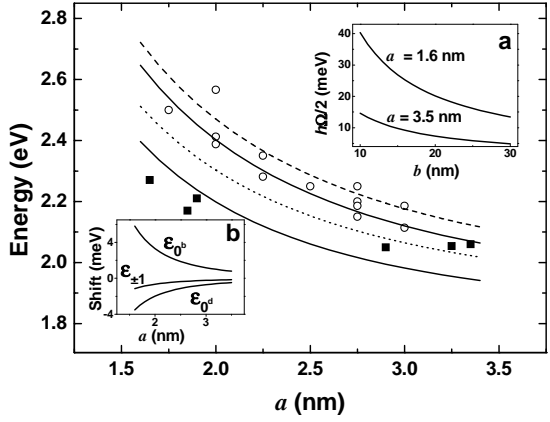, angle=-90, width=0.50\textwidth}
\end{center}
\vskip-0.5truecm
\caption{Size dependence of the transport and optical energy gap in CdSe NRs. Dotted and dashed lines show bare and dielectric confinement dressed energy between  the $1\Sigma_{1/2}$ hole and $1\Sigma_e$ electron subband. Two solid lines show the optical transition energy for the two first optically active  1DEs. Experimental data for the transport and optical gap measured in CdSe NRs from Ref. \cite{NRBanin} are shown by empty circles and filled squares. Insets a: Dependence of the 1DE localization energy, $\hbar\Omega/2$, on the NR semi-axis length $b$. b: Size dependence of the ground 1DE fine structure.} 
\vskip-0.1truecm
\label{polar}
\end{figure}

The above results for 1DEs in NRs can be applied directly to NWs. Unlike in NWs, however, the radius of NRs decreases and causes a rise of the electron and hole QSLs energy.  Both electrons and holes
tend to be at those parts of the NR where the radius is larger. In NRs of ideal ellipsoidal shape, this leads to the parabolic potential, acting on the 1DE center of mass motion, whose minimum is at the NR center and frequency is $\Omega = \sqrt{a[-\partial(E_{10} + E_{\Sigma_{1/2}})/\partial a]/b^2M}$, where $M = m(E)+m_{1\Sigma_{1/2}}$ is the 1DE mass. The inset in Fig.2 shows how the 1DE localization energy, $\hbar\Omega/2$,  depends on $b$ for two values of the NR radius. 

The PL properties of NRs are determined by the fine structure of the ground exciton state. The electron-hole  exchange interaction $\hat{H}_{\rm ex}= -(2\pi/3)\hbar\omega_{ST}(a_{\rm ex})^3\delta(\mbox{\boldmath$r$}_e-\mbox{\boldmath$r$}_h)(\mbox{\boldmath$s$}\cdot\mbox{\boldmath$J$})$ \cite{Nirmal}, where $\mbox{\boldmath$s$}$ is the electron spin 1/2 matrix, $a_{\rm ex}$ is the bulk exciton Bohr radius,  and  $\hbar\omega_{ST}$ is the bulk exciton singlet-triplet splitting (in CdSe: $\hbar\omega_{ST}=0.13$\,meV and $a_{\rm ex}=56$\,\AA), splits the four-fold degenerate ground state of the 1DE in NRs into three states: two states with total angular momentum projection $F_z=s_z+j_z=0$ and a degenerate state with $F_z=\pm 1$.   The calculation of the 1DE fine structure is similar to Ref. \cite{Nirmal} and it shows that the ground $0^{\rm d}$ 1DE is the optically passive state with $F_z=0$. The other two 1DE states, $\pm 1$ with $F_z=\pm 1$ and $0^{\rm b}$ with $F_z=0$, are optically active. The  size dependence of this fine structure is described as:
\be
\varepsilon_{0^{\rm d},0^{\rm b}} = \varepsilon_{\rm ex}[\eta(\beta) \mp \lambda(\beta)],~\varepsilon_{\pm1} = -\varepsilon_{\rm ex}\eta(\beta)
\ee
where $\varepsilon_{\rm ex}= \left(\pi/3\right)\hbar\omega_{ST}\left(a_{\rm ex}^3/a^2 \right)\Phi_{\rm ex}^2\left(0 \right)$ and $\eta(\beta)$, $\lambda(\beta)$ are the dimensionless functions of the ratio of the light and heavy hole effective masses $\beta$. In CdSe NR's $\eta(\beta)=0.34$ and $\lambda(\beta)=1.38$.  The dependence of the exciton fine structure on $a$, is shown in the inset (b) of Fig. 2.

For the exciton localized at the NR center, we have calculated the dipole matrix element  of the optical transition to the $0^{\rm b}$ exciton state: $d_{0^b}=\sqrt{2E_p/3m_0\omega^2}K_{1S,1S_{1/2}}(\pi\hbar/4M\Omega)^{1/4}\Phi_{\rm ex}(0)$, where $\omega$ is the optical transition frequency. This dipole is $2\sqrt{2}$ times larger than the $d_{\pm 1}$ dipoles of the optical transitions to the $\pm 1$ exciton states. In the ideal NR, the exciton PL intensity is proportional to the sum of the relative population of the $\pm 1$ and $0^{\rm b}$ exciton states multiplied by the corresponding matrix element squared. In addition, due to the NR anisotropy, the emission from the $0^b$ state is much stronger than from the $\pm 1$ states because the electric field of a photon is significantly reduced if the field is perpendicular to the NR axis while the photon field remains almost unchanged if it is parallel to the NR axis. This dielectric enhancement can be written as 
\begin{equation} 
  R_{\rm e} = \{[\kappa_m +(\kappa_s - \kappa_m)n^{(\bot)}]/[\kappa_m +(\kappa_s - \kappa_m)n^{(z)}]\}^2~,
\end{equation}
where  $n^{(z)} = a^2[\ln(2b/a) - 1]/b^2$ and $n^{(\bot)} = (1 - n^{(z)})/2$ are the depolarization coefficients for the two components
of an electric field parallel and perpendicular to the NR respectively. In the elongated ($b\gg a$) CdSe NRs $R_{\rm e}\approx [(\kappa_s+\kappa_m)/2\kappa_m]^2\approx 4.1$.  Taking into account the exciton level population, the anisotropy of the matrix elements, and the dielectric confinement, we have calculated the PL polarization degree of a single NR as $P=(I_\parallel-I_\perp)/(I_\parallel+I_\perp)$, where $I_\parallel$ and $I_\perp$ are the intensity of light with the polarization vector lying in the same plane with the NR axis and perpendicular to it \cite{NRpolarization}. The angular dependence of $P$ can be written:
\be
 P = \cos 2\psi[(1- x)\sin^2\theta ]/[\sin^2\theta + x(1 + \cos^2\theta)]~,
\ee
where $x = e^{(\varepsilon_{0^b} -\varepsilon_{\pm 1})/kT}/(4R_{\rm e})$, $T$ is the temperature, $\psi$ is the angle between the polarization analyzer and the NR axis, the observation angle $\theta$ is measured from a direction parallel to the NR. At room temperature and $\theta = \pi/2$,  the polarization is modulated at 87\%, which is in very good agreement with the experiment \cite{NRpolarization}. 

Our calculations show that the anisotropy of dielectric and spatial confinement  shortens the radiative lifetime  at room temperatures down to 500\, ps in reasonably elongated NRs  relative to $\sim 20$\,ns measured in CdSe NCs.  

The 1DE is localized at the NR center only if the NR thickness decreases to the edges. Any nonmonotonic variations of NR thickness, NR bending, and localized charges at the NR surface should lead to 1DE localization at different parts of the NR. This results in inhomogeneous broadening of 1DEs, even in a single NR.

To summarize, we have described the structure of the 1D excitons in NRs, which  control practically all aspects of the NR PL properties even at room temperatures. 

The authors acknowledge financial support from DARPA and ONR; A. S.  thanks the NRC for support.

\end{document}